\documentclass[aip,jmp,amsmath,amssymb,preprint]{revtex4-1}
\usepackage{graphicx}%
\usepackage{dcolumn}%
\usepackage{bm}%

\begin{document}

\newcommand{\pst}{\hspace*{1.5em}}
\newcommand{\be}{\begin{equation}}
\newcommand{\ee}{\end{equation}}
\newcommand{\ds}{\displaystyle}
\newcommand{\bdm}{\begin{displaymath}}
\newcommand{\edm}{\end{displaymath}}
\newcommand{\bea}{\begin{eqnarray}}
\newcommand{\eea}{\end{eqnarray}}

\title{Conditions for quantum and classical tomogram-like functions \\
to describe system states and to retain normalization during evolution}

\author{Ya. A. Korennoy, }
\author{V. I. Man'ko}
\affiliation{P.N. Lebedev Physical Institute,   \\
       Leninskii prospect 53, 119991, Moscow, Russia }

\date{4 August 2015}

\begin{abstract}\noindent
It is shown that dynamical equations for quantum tomograms retain 
the normalization conditions of their solutions during evolution only if the solutions 
satisfy a set of special conditions. These conditions are found explicitly.
On the contrary, it is also shown that the classical Liouville equation, 
Moyal  equation for  Wigner function, and evolution equation 
for Husimi function retain normalization 
of any initially normalized and quickly decaying at infinity functions on the phase space. 
Other necessary and sufficient conditions for optical and symplectic tomogram-like functions
to be tomograms of physical states are discussed.
\end{abstract}



\maketitle

\section{Introduction}

It was  found \cite{Mancini96} (see also \cite{IbortPhysScr}) that the quantum states can
be described by fair probability distributions, which contain the same complete
information of quantum states that is contained in density operator \cite{Landau1927}
and different its representations like Wigner function \cite{Wigner32},
Husimi Q-function \cite{Husimi40} and diagonal representation P-function
of Glauber \cite{Glauber63} and Sudarshan \cite{Sudarshan63}.
Such a probability distributions depend on homodyne variables and reference frames parameters
and are called  optical tomograms \cite{BerBer,VogRis} and symplectic tomograms
\cite{Mancini95}.

It is obvious, that not all of functions of these variables can be a tomogram
of a physical state of quantum or classical system. In Ref. \cite{OConnell}
two sets of conditions, which are necessary and sufficient for a function
defined on phase space to be a Wigner function, were discussed.
In Ref. \cite{Simoni2010} conditions for a symplectic tomogram-like function to be 
a tomogram were formulated based on association of symplectic tomograms 
with a unitary representation of the Weyl-Heisenberg group.

The evolution equation of symplectic tomograms of quantum systems was obtained in \cite{ManciniFoundPhys97}.
For the optical tomograms such a dynamical equation was firstly  found in \cite{Korarticle2}.
We also tried to obtain the tomographic evolution equations for relativistic
systems in Ref. \cite{Korarticle1}. 

The aim of our article is to explore the conditions, which must be satisfied by
optical tomograms of both quantum and classical states. 
We focus on the problem of the conditions for optical tomograms due to these
tomograms are directly measured by homodyne detectors in quantum optical
experiments (see \cite{Raymer93,LvovRayRevModPhys}).

We also study the conditions under which the tomographic evolution equations 
retain the normalization of their solutions.
It turns out that these conditions play a decisive role in numerical findings of the solutions
of such an equations because these equations conserves normalizations only 
for a narrow class of functions, which obey to the special relations,
and limited accuracy of numerical calculations can lead to non-normalized solutions.

The paper is organized as follows. In Sec.\ref{Section2} we review the definitions and main properties of optical
and symplectic tomograms for both classical and quantum states.
In Sec.\ref{Section3} conditions for  optical and symplectic tomogram-like functions  to be 
tomograms of a physical states are given.
In Sec.\ref{Section4} conditions for optical tomogram-like functions to retain normalization
during evolution are obtained.
In Sec.\ref{Section5} conditions for conservation of 
normalization of the symplectic tomograms during evolution are found.
In Sec.\ref{Section6} we show that Liuoville equation, dynamical equations for Wigner function and Husimi function
do not demand of any additional conditions to conserve
normalizations of their solutions.
In Sec.\ref{Section7} conclusion and prospectives are presented.

\section{\label{Section2}Definitions and properties of optical and symplectic tomograms}

For spinless $N$ dimensional quantum systems  with the density matrix $\hat \rho(t)$, which can be 
dependent on time,  the corresponding 
optical $w(\mathbf X,\bm\theta,t)$ and symplectic $M(\mathbf X,\bm\mu,\bm\nu,t)$ tomograms are defined as follows
(see \cite{Korarticle5,IbortPhysScr})
\be		\label{eq_43}
w(\mathbf X,\bm\theta,t)=\mathrm{Tr}\left\{
\hat\rho(t) \prod_{\sigma=1}^{N}
\delta \left(X_\sigma-\hat q_\sigma\cos\theta_\sigma
-\hat p_\sigma\frac{\sin\theta_\sigma}{m_\sigma\omega_\sigma}\right)
\right\},
\ee
\be		\label{eq_44}
M(\mathbf X,\bm\mu,\bm\nu,t)=\mathrm{Tr}\left\{
\hat\rho(t) \prod_{\sigma=1}^{N}
\delta \left(X_\sigma-\mu_\sigma \hat q_\sigma-\nu_\sigma \hat p_\sigma\right)
\right\},
\ee
where frequency $\omega_{\sigma}$ and mass $m_\sigma$ dimensional constants 
for $\sigma-$th degree of freedom  are chosen from convenience considerations for particular Hamiltonian.
To simplify the formulas hereafter we choose the system of units
such that $m_\sigma=\omega_\sigma=\hbar=1$.
Given the Wigner function of the state of quantum system or the classical
distribution function on phase space $W(\mathbf q,\mathbf p,t)$, for optical and symplectic tomograms
we can write
\be		\label{optfromWig}
w(\mathbf X,\bm\theta,t)=
\int~W(\mathbf q,\mathbf p,t)\prod_{\sigma=1}^{N}
\delta \left(X_\sigma-q_\sigma\cos\theta_\sigma
-p_\sigma\sin\theta_\sigma\right)
d^Nq~d^Np,
\ee
\be		\label{sympfromWig}
M(\mathbf X,\bm\mu,\bm\nu,t)=
\int~W(\mathbf q,\mathbf p,t)\prod_{\sigma=1}^{N}
\delta (X_\sigma-\mu_\sigma q_\sigma-\nu_\sigma p_\sigma)d^Nq~d^Np.
\ee
We suppose that  the normalization of the Wigner function is unity.

The inverse transforms of maps (\ref{optfromWig}) and (\ref{sympfromWig})
are given by the Fourier integrals: 
\be		\label{eq_52}
W(\mathbf q,\mathbf p,t)=\int\limits_{0}^{\pi}d^N\theta
\int\limits_{-\infty}^{+\infty}\frac{d^N\eta~d^NX}{(2\pi)^{2N}}
w(\mathbf X,\bm\theta,t)\prod_{\sigma=1}^{N}
|\eta_\sigma|\exp\left\{i\eta_\sigma\left(X_\sigma
-q_\sigma\cos\theta_\sigma-p_\sigma\sin\theta_\sigma\right)\right\} ,
\ee
\be		\label{eq_53}
W(\mathbf q,\mathbf p,t)=\frac{1}{(2\pi)^{2N}}\int M(\mathbf X,\bm\mu,\bm\nu,t)
\prod_{\sigma=1}^{N}\exp\left\{ i \left(X_\sigma-\mu_\sigma q_\sigma
-\nu_\sigma p_\sigma\right)\right\}d^NX~d^N\mu~d^N\nu.
\ee

Tomograms  are nonnegative and normalized by the conditions
\be				\label{eqnormOpt}
\int w(\mathbf X,\bm\theta,t)d^NX=1,
\ee
\be				\label{eqnormSymp}
\int M(\mathbf X,\bm\mu,\bm\nu,t)d^NX=1.
\ee

Since the Dirac delta-function in Eq.(\ref{eq_44}) is homogeneous function,
i.e. $\delta(\lambda y)=|\lambda|^{-1}\delta(y)$, the symplectic tomogram is also the homogeneous function
\be                             \label{eq_3}
M(\lambda_\sigma X_\sigma,\lambda_\sigma\mu_\sigma,\lambda_\sigma\nu_\sigma,t)=
M(\mathbf X,\bm\mu,\bm\nu,t)\prod_{\sigma=1}^N|\lambda_\sigma |^{-1}.
\ee
The optical tomogram is  even function. It means that the optical tomogram has the property
\be		\label{eq_14}
w(-X_\sigma,\theta_\sigma+\pi,t)=w(\mathbf X,\bm\theta,t).
\ee
It is obvious, that 
\be		\label{OptSym}
w(\mathbf X,\bm\theta,t)=M(\mathbf X,\mu_\sigma=\cos\theta_\sigma,\nu_\sigma=\sin\theta_\sigma,t).
\ee
The homogeneity condition (\ref{eq_3}) provides the relation of the optical and symplectic tomograms opposite to (\ref{OptSym}).
One can transform the integral (\ref{sympfromWig}) using polar coordinates 
$\mu_\sigma=r_\sigma\cos\theta_\sigma$, $\nu_\sigma=r_\sigma\sin\theta_\sigma$. After some calculations
for $N-$dinentional case we can write
\be                             \label{eq_6}
M(\mathbf X,\bm\mu,\bm\nu)=w\left(\frac{X_\sigma\mathrm{sgn}(\nu_\sigma)}{\sqrt{\mu_\sigma^2+\nu_\sigma^2}},
\cot^{-1}\frac{\mu_\sigma}{\nu_\sigma}\right)
\prod_{\sigma=1}^N\frac{1}{\sqrt{\mu_\sigma^2+\nu_\sigma^2}}.
\ee
Also the symplectic tomogram satisfy an extra homogeneous
differential relation. Making differentiation of (\ref{eq_3}) by $\lambda_i$
and substituting $\lambda_i=1$ we obtain
$$
M+X_i\partial_{X_i}M+\mu_i \partial_{\mu_i}M
+\nu_i\partial_{\nu_i}M=0.
$$
In addition, optical tomograms must satisfy the entropic uncertainty relation
(see \cite{MAMankoAIP2011}) (Hirshman criterion)
\be			\label{Hirschman}
-\int w(\mathbf X,\bm\theta,t)\ln w(\mathbf X,\bm\theta,t)d^NX
-\int w(\mathbf X,\bm\theta+\bm{\pi}/2,t)\ln w(\mathbf X,\bm\theta+\bm{\pi}/2,t)d^NX \geq N\ln(\pi e).
\ee
Here $S(\bm\theta,t)=-\int w(\mathbf X,\bm\theta,t)\ln w(\mathbf X,\bm\theta,t)d^NX$
is the tomographic Shannon entropy associated with optical tomogram.

In Ref. \cite{Simoni2010} it was mentioned that non-negativity, normalization and homogeneity
properties are not sufficient for the symplectic tomogram-like function $f(\mathbf X,\bm\mu,\bm\nu)$
to be a tomogram. Also non-negativity, normalization (\ref{eqnormOpt}), parity (\ref{eq_14}),
and satisfaction of Hirshman criterion (\ref{Hirschman}) are not sufficient for optical 
tomogram-like function $f(\mathbf X,\bm\theta)$ to be a tomogram of any quantum or
classical state.
For example, consider
\be		\label{example1}
f(X,\theta)=\frac{1}{\sqrt\pi}e^{-(X-\cos^3\theta)^2}.
\ee
In spite of this function is positive and satisfy  conditions (\ref{eqnormOpt}),
(\ref{eq_14}), and (\ref{Hirschman}), it is not an optical tomogram. More over, as it will
be seen from Section \ref{Section4}, the normalization of this function will not be conserved 
during time evolution.

\section{\label{Section3}Conditions for tomogram-like functions  to be 
tomograms of physical states}

In Ref. \cite{OConnell} necessary and sufficient conditions for a phase-space function 
to be a Wigner distribution of physical system are studied. Such function is a Wigner function
if and only if it defines a kernel (density matrix) for a positive trace-class operator
with trace equal to one. The authors consider two sets of necessary and
sufficient conditions and show that these sets are formally equivalent.

\textbf{The first} more familiar \textbf{set} is that the function $W(\mathbf q,\mathbf p)$ must be
square integrable, and for every Wigner function $W_\psi(\mathbf q,\mathbf p)$
of a pure state $\psi$ the following inequality must be valid
\be		\label{conditionW1}
\int W(\mathbf q,\mathbf p)W_\psi(\mathbf q,\mathbf p)d^Nq~d^Np\geq 0,
\ee
and the function $W(\mathbf q,\mathbf p)$ must be normalized to unity
over all phase space,
\be		\label{normWig}
\int W(\mathbf q,\mathbf p)d^Nq~d^Np=1.
\ee
One may find the proof that these conditions are necessary and
sufficient in \cite{PhisRep1984}.

Let us note that from the physical point of view the integral in inequality 
(\ref{conditionW1}) is a transition probability from the state with Wigner function 
$W(\mathbf q,\mathbf p)$ to the state $\psi$ and, obviously, it must be non-negative.

\textbf{The second set} of conditions is called in \cite{OConnell} as KLM conditions,
after Kastler \cite{Kastler}, Loupias, and Miracle-Sole \cite{Loupias1,Loupias2},
and applied to the symplectic Fourier transform of the 
phase space function $W(\mathbf q,\mathbf p)$
\be		\label{sympFour}
\widetilde W(\mathbf{u},\mathbf{v})=\int W(\mathbf{q},\mathbf{p})
e^{i(\mathbf q\mathbf v-\mathbf u\mathbf p)}d^Nq~d^Np.
\ee
The first KLM condition, which is equivalent to (\ref{conditionW1}), is that the function 
$\widetilde W(\mathbf{u},\mathbf{v})$ must be continuous and of $\hbar-$positive type,
i.e. for every choce of poins 
$(\mathbf u_1,\mathbf v_1)$, $(\mathbf u_2,\mathbf v_2)$, ..., $(\mathbf u_n,\mathbf v_n)$,
the $n\times n$ matrix  with entries 
\be		\label{matrixnonneg1}
Z_{jk}=e^{i(\mathbf u_j\mathbf v_k-\mathbf u_k\mathbf v_j)/2}
\widetilde W(\mathbf u_j-\mathbf u_k,\mathbf v_j-\mathbf v_k)
\ee
is non-negative (in our system of units $\hbar=1$).
The second KLM condition is equivalent to (\ref{normWig})
\be		\label{normWig2}
\widetilde W(\mathbf u=0,\mathbf v=0)=1.
\ee

Using the formulas of maps of the Wigner function  to the tomogram and vice versa
one can find necessary and sufficient conditions for tomographic functions corresponding to (\ref{conditionW1}).
Normalized according to (\ref{eqnormOpt}) and (\ref{eqnormSymp}) functions $w(\mathbf X,\bm\theta)$ and $M(\mathbf X,\bm\mu,\bm\nu)$ 
obeying relations  (\ref{eq_14}) or (\ref{eq_3})  are tomograms of
states of quantum systems if and only if for any quantum pure state $\psi$ with
the tomogram $w_\psi(\mathbf X,\bm\theta)$ (or $M_\psi(\mathbf X,\bm\mu,\bm\nu)$),
the following inequalities are valid
\be			\label{conditwpos}
\int w(\mathbf X,\bm\theta)w_\psi(\mathbf X,\bm\theta)
e^{i\bm\eta(\mathbf X-\mathbf X')}|\eta_1||\eta_2|\dots|\eta_n|
d^nX~d^NX'~d^n\eta~d^N\theta\geq 0, 
\ee
\be			\label{conditMpos}
\int M(\mathbf X,\bm\mu,\bm\nu)M_\psi(\mathbf X',\bm\mu,\bm\nu)
e^{i(\mathbf X-\mathbf X')}
d^nX~d^nX'~d^N\mu~d^N\nu\geq 0. 
\ee
Also with the map between functions $W(\mathbf q,\mathbf p,t)$ and $M(\mathbf X,\bm\mu,\bm\nu,t)$
 the symplectic Fourier transform (\ref{sympFour}) is expressed as follows
\be		\label{symFsym}
\widetilde W(\mathbf{u},\mathbf{v})=\int M(\mathbf X,\mathbf{v},-\mathbf{u})
e^{i\mathbf X}d^NX.
\ee
Substituting (\ref{symFsym}) to the first KLM condition with the set of points
$\{(\bm\mu_j,\bm\nu_j):\mathbf{v}_j=\bm\mu_j,\mathbf{u}_j=-\bm\nu_j \}$,
which will also be arbitrary, we obtain the expression for non-negative matrix $Z_{jk}$
from symplectic tomogram
\be		\label{matrixnonneg2}
Z_{jk}=e^{i(\bm\mu_j\bm\nu_k-\bm\mu_k\bm\nu_j)/2}
\int M(\mathbf X,\bm\mu_j-\bm\mu_k,\bm\nu_j-\bm\nu_k)e^{i\mathbf X}d^NX.
\ee
With the help of relation (\ref{eq_6}) between $M(\mathbf X,\bm\mu,\bm\nu)$ and
$w(\mathbf X,\bm\theta)$ we immediately obtain the expression for the matrix $Z_{jk}$
in terms of the optical tomogram
\bea		
Z_{jk}&=&e^{i(\bm\mu_j\bm\nu_k-\bm\mu_k\bm\nu_j)/2} \nonumber \\[3mm]
&&\times\int w\left(X_\sigma,
\cot^{-1}\frac{\mu_{\sigma j}-\mu_{\sigma k}}{\nu_{\sigma j}-\nu_{\sigma k}}\right)
\nonumber \\[3mm]
&&\times\exp\left[i\sum_{\sigma=1}^N X_\sigma\mathrm{sgn}(\nu_{\sigma j}-\nu_{\sigma k})
\sqrt{(\mu_{\sigma j}-\mu_{\sigma k})^2+(\nu_{\sigma j}-\nu_{\sigma k})^2}\right]d^NX.
\label{matrixnonneg3}
\eea
As for condition (\ref{normWig2}), it will be valid as the limit case 
\be		\label{normWig3}
\widetilde W(0,0)=\int M(\mathbf X,0,0)e^{i\mathbf X}d^NX=
\int w\left(\mathbf X ,\cot^{-1}\frac{0}{0}\right)d^NX=1,
\ee
because
the symplectic tomogram transforms to the delta-function
$M(\mathbf X,0,0)=\delta(\mathbf X)$, and the function $\cot^{-1}(x)$ is confined at any argument.
Thus, instead of (\ref{normWig3}) it is more preferably to use conditions (\ref{eqnormOpt}) and (\ref{eqnormSymp}),
which must be valid for any $(\bm\mu,\bm\nu)\neq0$, or phase vector $\bm\theta$ $\{\theta_j\in[0,\pi]\}$.

In Ref. \cite{Simoni2010} the same expression (\ref{matrixnonneg2}) 
for non-negative matrix $Z_{jk}$ was obtained from considerations of group theory
and called as $\omega-$positivity condition. It was shown \cite{Simoni2009} that symplectic tomograms 
are associated with nontrivial unitary irreducible representations
(we generalize here the results of \cite{Simoni2010,Simoni2009} to $N-$dimensional case
and to optical tomograms)
\be		\label{Weilrep}
\hat U_g(\bm\mu,\bm\nu,\xi)=\exp[i(\bm\mu\hat{\mathbf q}+\bm\nu\hat{\mathbf p}\,)]e^{i\xi}
\ee
of the Weyl-Heisenberg group WH(2N). So, according to Naimark's theorem
\cite{Naimark} the function 
\bea		
\varphi(\bm\mu,\bm\nu,\xi)&=&\mathrm{Tr}\left\{\hat\rho
\exp\left[i\left(\bm\mu\hat{\mathbf q}+\bm\nu\hat{\mathbf p}\,\right)\right]\right\}e^{i\xi}
=e^{i\xi}\int M(\mathbf X,\bm\mu,\bm\nu)e^{i\mathbf X}d^NX \nonumber \\[3mm]
&=&e^{i\xi}\int w\left(\mathbf X,\cot^{-1}(\mu_\sigma/\nu_\sigma)\right) 
\exp\left[i\sum_{\sigma=1}^N X_\sigma\mathrm{sgn}(\nu_{\sigma})
\sqrt{\mu_{\sigma}^2+\nu_{\sigma}^2}\right]d^NX
\label{funcongr}
\eea
is a positive definite function on the Weyl-Heisenberg group, i.e.
for any $n-$tuple of group elements $(g_1,g_2,...,g_n)\in \mathrm{WH}(2N)$
the $n\times n$ matrix with elements
\be		\label{matrixnonneg4}
Z_{jk}=\varphi(g_jg_k^{-1})
\ee
is non-negative. Substitution of the expression of representation of the 
group element $g_jg_k^{-1}$
\be		\label{prod1}
\hat U_{g_j}(\bm\mu_j,\bm\nu_j,\xi_j)\hat U_{g_k}^{-1}(\bm\mu_k,\bm\nu_k,\xi_k)=
\hat U_{g_jg_k^{-1}}\Big[\bm\mu_j-\bm\mu_k,\bm\nu_j-\bm\nu_k,\xi_j-\xi_k+
(\bm\mu_j\bm\nu_k-\bm\mu_k\bm\nu_j)/2\Big]
\ee
to  definition (\ref{funcongr}) of the function $\varphi$ and omission of the 
inessential factors $e^{i\xi_j}$ and $e^{i\xi_k}$ lead us to expressions of the 
matrix $Z_{jk}$ for symplectic (\ref{matrixnonneg2}) and optical (\ref{matrixnonneg3})
tomograms.

In order to functions $w_\mathrm{cl}(\mathbf X,\bm\theta)$ or $M_\mathrm{cl}(\mathbf X,\bm\mu,\bm\nu)$
be tomograms of classical systems, in addition to their smoothness, normalization in accordance with 
conditions (\ref{eqnormOpt}, \ref{eqnormSymp}), parity (\ref{eq_14}) or homogeneity (\ref{eq_3}) respectively,
necessary and sufficient conditions is that they must define the function $W_\mathrm{cl}(\mathbf q,\mathbf p)$ 
with formula (\ref{eq_52}) or (\ref{eq_53}), which consist in a class of distribution functions,
i.e. positive normalized functions.

In terms of group theory  the function $\phi(\bm\mu,\bm\nu)$ 
with the definition
\bea		
\phi(\bm\mu,\bm\nu)&=&\int W_\mathrm{cl}(\mathbf q,\mathbf p)
\exp\left[i\left(\bm\mu\mathbf{q}+\bm\nu\mathbf{p}\,\right)\right]d^Nq~d^Np
=\int M_\mathrm{cl}(\mathbf X,\bm\mu,\bm\nu)e^{i\mathbf X}d^NX \nonumber \\[3mm]
&=&\int w_\mathrm{cl}\left(\mathbf X,\cot^{-1}(\mu_\sigma/\nu_\sigma)\right) 
\exp\left[i\sum_{\sigma=1}^N X_\sigma\mathrm{sgn}(\nu_{\sigma})
\sqrt{\mu_{\sigma}^2+\nu_{\sigma}^2}\right]d^NX
\label{funcongrcl}
\eea
on the translational group with group law 
\be			\label{transgrlaw}
(\bm\mu,\bm\nu)\circ(\bm\mu\,',\bm\nu\,')=(\bm\mu+\bm\mu\,',\bm\nu+\bm\nu\,')
\ee 
must be a positive definite function on this group (see \cite{Simoni2010}), 
i.e. for every choice of points
$(\bm\mu_1,\bm\nu_1)$, $(\bm\mu_2,\bm\nu_2)$, ..., $(\bm\mu_n,\bm\nu_n)$
the $n\times n$ matrix  with entries $\phi_{jk}=\phi(\bm\mu_j-\bm\mu_k,\bm\nu_j-\bm\nu_k)$
must be non-negative.
Then, by the Bochner theorem the function $\phi(\bm\mu,\bm\nu)$ is the Fourier transform 
of the probability measure on the phase space.

Note \cite{Simoni2010}  that the function $\phi(\bm\mu,\bm\nu)=e^{-i\xi}\varphi(\bm\mu,\bm\nu,\xi)$
may be simultaneously of $\hbar-$positive and positive type on the translational group
with group law (\ref{transgrlaw}).
 In this case the corresponding tomogram-like function 
$w(\mathbf X,\bm\theta)$ or $M(\mathbf X,\bm\mu,\bm\nu)$ may be interpreted as a quantum as a classical tomogram.
Such an example is the tomogram of the ground state of the harmonic oscillator.

Let's additionally note that conditions considered are always necessary but they are sufficient if and only if 
the Radon integral  of tomogram-like function of type (\ref{optfromWig})  
or (\ref{sympfromWig}) with $W(\mathbf q,\mathbf p,t)$
defined by (\ref{eq_52}) or (\ref{eq_53})  exist and equal to the function itself.
As an example, examine the function
\be			\label{exampfunc1}
f_1(X,\mu,\nu)=\frac{e^{1/4}}{\sqrt\pi}e^{-X^2-\mu^2/4-\nu^2/4}.
\ee
Evidently, that this function is neither quantum no classical tomogram.
Nevertheless, it is easy to see that both the first and the second KLM conditions for this
function are satisfied:
\bdm
\widetilde f_1(\mu,\nu)=\int f_1(X,\mu,\nu)e^{iX}dX=
e^{-\mu^2/4-\nu^2/4},~~~~\widetilde f_1(X,0,0)=1,
\edm
and $\widetilde f_1(\mu,\nu)$ is $\hbar-$positive function.
The thing is that, if we apply transformation (\ref{eq_53}) to this function, 
we obtain the classical distribution $W_1(q,p)=\pi^{-1/2}e^{-q^2-p^2}$ 
(which in this special example  can also be interpreted as
a Wigner function of  the ground state of the harmonic oscillator),
but if we apply to $W_1(q,p)$ map (\ref{sympfromWig}), which is inverse to (\ref{eq_53}), we  obtain 
the function that does not equal to $f_1(X,\mu,\nu)$.
So, Radon integral (\ref{eq_53})  of the function $f_1(X,\mu,\nu)$ exist but does not 
equal to the function $f_1(X,\mu,\nu)$, and KLM conditions are not  sufficient 
for the tomogram-like function $f_1(X,\mu,\nu)$ to be a tomogram of the state of a physical quantum 
or classical system.

\section{\label{Section4}Conditions for conservation of normalization \\ 
of optical tomogram-like functions during evolution}
For Hamiltonians
\be                             \label{r18_2}
\hat H=\sum_\sigma\frac{\hat p_\sigma^2}{2m_\sigma}+V({\mathbf q},t)
\ee
the evolution equation 
of the optical tomogram  $w({\mathbf X},{\bm\theta},t)$ 
firstly was found in \cite{Korarticle2}
\begin{eqnarray}                             
&&
\frac{\partial}{\partial t}w({\mathbf X},{\bm\theta},t)=
\sum_{\sigma=1}^N\omega_{0\sigma}
\left[\cos^2\theta_\sigma\frac{\partial}{\partial\theta_\sigma}
-\frac{1}{2}\sin2\theta_\sigma\left\{1+X_\sigma\frac{\partial}{\partial X_\sigma}\right\}
\right]
w({\mathbf X},{\bm\theta},t)\nonumber \\[3mm]
&&+
\frac{2}{\hbar}\left[\mathrm{Im}~V\left\{
\sin\theta_\sigma\frac{\partial}{\partial\theta_\sigma}
\left[\frac{\partial}{\partial X_\sigma}\right]^{-1}
+X_\sigma\cos\theta_\sigma+i\frac{\hbar\sin\theta_\sigma}
{2m_\sigma\omega_{0\sigma}}
\frac{\partial}{\partial X_\sigma}\right\}\right]
w({\mathbf X},{\bm\theta},t), 
\label{r20_2}
\end{eqnarray}
where we introduced the designation \cite{Korarticle5}
\be			\label{invder}
\left[\frac{\partial}{\partial X_\sigma}\right]^{-n}f(X_\sigma)=\frac{1}{(n-1)!}
\int(X_\sigma-X_\sigma')^{n-1}\Theta(X_\sigma-X_\sigma')f(X_\sigma')dX_\sigma',
\ee
where $\Theta(X_\sigma-X_\sigma')$ is a Heaviside step function.
Optical tomographic representation of the classical non-relativistic Liouville equation
\cite{Korarticle2} is the limit case of (\ref{r20_2}) when $\hbar\to 0$
\begin{eqnarray}                             
&&\ds{\frac{\partial}{\partial t}w_\mathrm{cl}(\mathbf X,\bm\theta,t)=
\sum_{\sigma=1}^{N}\omega_\sigma \left[\cos^2\theta_\sigma\frac{\partial}{\partial\theta_\sigma}
-\frac{1}{2}\sin2\theta_\sigma\left\{1+X_\sigma\frac{\partial}{\partial X_\sigma}\right\}
\right]w_\mathrm{cl}(\mathbf X,\bm\theta,t)} \nonumber \\[3mm]
&&\ds{+\left[\sum_{\sigma=1}^{n}\frac{\partial}{\partial q_\sigma}~V\left\{
q_\sigma\rightarrow\sin\theta_\sigma\frac{\partial}{\partial\theta_\sigma}
\left[\frac{\partial}{\partial X_\sigma}\right]^{-1}
+X_\sigma\cos\theta_\sigma\right\}
\frac{\sin\theta_\sigma}{m_\sigma\omega_\sigma}\frac{\partial}{\partial X_\sigma}\right]
w_\mathrm{cl}(\mathbf X,\bm\theta,t).}
		\label{eq_54}
\end{eqnarray}
Evolution equation of the  tomogram for arbitrary spinless
quantum Hamiltonian  was found in \cite{Korarticle1}.

Equations (\ref{r20_2}), (\ref{eq_54}) can be presented in the following compact form
\be 			\label{eqcompFF}
\partial_t w({\mathbf X},{\bm\theta},t)=\hat F_\mathrm{k}({\mathbf X},{\bm\theta})~w({\mathbf X},{\bm\theta},t)
+\hat F_\mathrm{p}({\mathbf X},{\bm\theta},t)~w({\mathbf X},{\bm\theta},t),
\ee
where $\hat F_\mathrm{k}({\mathbf X},{\bm\theta})$ is a time independent operator
corresponding to the kinetic part of the Hamiltonian, and  
$\hat F_\mathrm{p}({\mathbf X},{\bm\theta},t)$ is an operator
corresponding to the potential energy of the system. 
Generally speaking, $\hat F_\mathrm{p}({\mathbf X},{\bm\theta},t)$ can be dependent on time
if the potential $V(\mathbf q,t)$ depend on time. More over, the operator 
$\hat F_\mathrm{k}$ for equations (\ref{r20_2}) and (\ref{eq_54}) is the same,
but the operators $\hat F_\mathrm{p}$ are different for the classical and quantum cases.

If we know the initial condition $w_0({\mathbf X},{\bm\theta})$
for the equation with the form (\ref{eqcompFF}),
we can write its formal general solution with the help of chronological exponential operator 
\be			\label{solexp}
w({\mathbf X},{\bm\theta},t)=\mathrm{T}\exp\left\{
\int\limits_0^t\left[\hat F_\mathrm{k}({\mathbf X},{\bm\theta})+ 
\hat F_\mathrm{p}({\mathbf X},{\bm\theta},t)\right]dt\right\}w_0({\mathbf X},{\bm\theta}).
\ee 

Which of the properties the tomograms must satisfy for conservation
of the normalization  during evolution?
Let's integrate left and right sides of equation (\ref{eqcompFF})
over the hyperspace $X^N$. If $w({\mathbf X},{\bm\theta},t)$ is a continuous
function normalized by unity
and tends to zero faster then any finite negative power of $X_i$ when $X_i$
tends to infinity, then the following relation must be valid
\be 			\label{intzero}
\partial_t \int w({\mathbf X},{\bm\theta},t)dX^N=0=
\int\hat F_\mathrm{k}({\mathbf X},{\bm\theta})~w({\mathbf X},{\bm\theta},t)dX^N
+\int\hat F_\mathrm{p}({\mathbf X},{\bm\theta},t)~w({\mathbf X},{\bm\theta},t)dX^N.
\ee
Using integration by parts and taking into account condition (\ref{eqnormOpt})
it is easy to show that 
\bdm
\int\hat F_\mathrm{k}({\mathbf X},{\bm\theta})~w({\mathbf X},{\bm\theta},t)d^NX=0.
\edm
For free motion $\hat F_\mathrm{p}\equiv 0$ and equation (\ref{intzero}) is valid
without any additional properties of the tomogram. For linear potential
$V(\mathbf q)$ equation (\ref{intzero}) is valid too.
For quadratic potential equation (\ref{eqcompFF}) transforms to 
\be			\label{eqossil}
\frac{\partial}{\partial t} w({\mathbf X},{\bm\theta},t)=\sum_{\sigma=1}^{N} \omega_\sigma
\frac{\partial}{\partial\theta_\sigma}w({\mathbf X},{\bm\theta},t),
\ee
and we see that there is only normalization property (\ref{eqnormOpt})
must be satisfied for tomograms to be normalized during evolution, i.e. for validity of equation (\ref{intzero}).

Further, to simplify the formulas, we consider one-dimensional motion and choose a system 
of physical units in which $\hbar=m=\omega_0=1$. The generalization to multidimensional case is straightforward.

As we discuss the related motion, then without loss of physical 
generality, it suffices to consider the case when $V(q)$ is 
a polynomial, because according to Chebyshev theorem, every continuous function $V(q)$ 
on a finite interval with any accuracy can be approximated by a polynomial.
Each term of the $n$-th power of $q$ of the polynomial potential will give 
contribution to (\ref{intzero}) as a sum of terms proportional to
$\cos^k\theta\sin^l\theta,$ where $k+l=n$.

Since equation (\ref{intzero}) must be satisfied for any fixed 
$\theta$ and $t$, then the sum of the coefficients of each term of the form 
$\cos^k\theta\sin^l\theta$ must be zero. Hence, we obtain 
a system of equations for tomogram, which  must be satisfied 
for any fixed $\theta$ and $t$.

Suppose that the potential $V(q)$ contains a term with $q^3$. 
Then, after some calculations, we can show that the quantum case 
equation (\ref{intzero}) will have the form
\begin{eqnarray}                             
0&=&-\frac{\sin^3\theta}{4}\int\partial _Xw(X,\theta,t)dX
\nonumber \\[3mm]
&&+3\sin\theta\cos^2\theta\left(2\int Xw(X,\theta,t)dX
+\int X^2\partial _Xw(X,\theta,t)dX
\right)
\nonumber \\[3mm]
&&+6\sin^2\theta\cos\theta~\partial_\theta\left(
\int \partial^{-1}_Xw(X,\theta,t)dX
+\int Xw(X,\theta,t)dX
\right)
\nonumber \\[3mm]
&&+3\sin^3\theta\left(
\partial^2_\theta \int\partial^{-1}_Xw(X,\theta,t)dX
-\int Xw(X,\theta,t)dX
\right).
\label{Vq3}
\end{eqnarray}
The same equation but without term in the first row is obtained for a classical motion. 
Using integration by parts it is easy to show that 
each of the first three rows in equation (\ref{Vq3}) is equal to zero, 
and fourth line leads us to the equation
\be			\label{ostatok}
\partial^2_\theta \int Xw(X,\theta,t)dX
+\int Xw(X,\theta,t)dX=0.
\ee
It is easy to see that this is a classical harmonic oscillator equation with variable $\theta$, 
and it is performed only if
\be			\label{garmsol}
\int Xw(X,\theta,t)dX=A(t)\cos\theta+B(t)\sin\theta.
\ee
Thus, we have the integral condition, which must be satisfied 
for the optical tomogram both in the classical and in the quantum case 
in order to its normalization will be remained during evolution in the third degree polynomial potential.
Note that from definition of tomogram (\ref{eq_43}) 
we can see that the left side of equality (\ref{garmsol}) is a mean value of the
homodyne variable $X$,
\be			\label{garmsolQP}
\int Xw(X,\theta,t)dX=\langle\hat q\rangle\cos\theta
+\langle\hat p\rangle\sin\theta,
\ee
where $\langle\hat q\rangle$ and $\langle\hat p\rangle$ are average quantum (classical) 
values of the position and momentum respectively.
So, we have $A(t)=\langle q\rangle$ and $B(t)=\langle p\rangle$.

Suppose that the potential $V(q)$ contains a term with $q^4$. 
After similar calculations, taking into account (\ref{eqnormOpt}) and (\ref{ostatok})
we obtain the differential equation for the second moment 
variable $X$, which is valid both in the quantum and in the classical case:
\be			\label{ostatok2}
\partial^3_\theta \int X^2w(X,\theta,t)dX
+4\partial_\theta \int X^2w(X,\theta,t)dX=0.
\ee
The solution of (\ref{ostatok2}) contains three constants independent on $\theta$, but dependent on time 
\be			\label{garmsol2}
\int X^2w(X,\theta,t)dX=A_1(t)\cos2\theta+B_1(t)\sin2\theta +C_1(t).
\ee
From definition of tomogram  (\ref{eq_43}) and the expression for the 
mean value of $X^2$ we can find the constants
$A_1(t)$, $B_1(t)$, $C_1(t)$,
\be			\label{garmsolQP2}
\int X^2w(X,\theta,t)dX=\langle\hat q^2\rangle\cos^2\theta
+\langle\hat p^2\rangle\sin^2\theta
+\langle\hat q\hat p+\hat p\hat q\rangle\sin\theta\cos\theta,
\ee
\bdm
A_1(t)=\frac{\langle\hat q^2\rangle-\langle\hat p^2\rangle}{2},~~~
B_1(t)=\frac{\langle\hat q\hat p+\hat p\hat q\rangle}{2},
~~~ C_1(t)=\frac{\langle\hat q^2\rangle+\langle\hat p^2\rangle}{2}.
\edm
Expressing  condition (\ref{garmsol2}) - (\ref{garmsolQP2}) through the Wigner function we obtain
\bea			
\int X^2w(X,\theta,t)dX&=&
\left(\int q^2W(q,p,t)dq~dp\right)\cos^2\theta
+\left(\int p^2W(q,p,t)dq~dp\right)\sin^2\theta \nonumber \\[3mm]
&&+2\left(\int qp\,W(q,p,t)dq~dp\right)\sin\theta\cos\theta.
\label{garmsolQP2W}
\eea
Arguing similarly, after some calculations we can show that if $V(q)$ contains a term with $q^n$, 
then all average values $\langle X^m\rangle$ of the variable $X$, where $m=1,~2,...,n-2$, 
must satisfy one of the following differential equation:
\be			\label{Gendifeven}
\left\{\frac{\partial}{\partial\theta}\prod_{k=1}^{m/2}
\left[\frac{\partial^2}{\partial\theta^2}+(2k)^2\right]\right\}
\int X^{m}\,w(X,\theta,t)dX=0,
\ee
if $m$ is even number, or 
\be			\label{Gendifodd}
\left\{\prod_{k=0}^{(m-1)/2}
\left[\frac{\partial^2}{\partial\theta^2}+(2k+1)^2\right]\right\}
\int X^{m}\,w(X,\theta,t)dX=0,
\ee
if $m$ is odd number.

General solutions of these equations are respectively equal:
\be			\label{Gensoleven}
\int X^{m}\,w(X,\theta,t)dX=A_0(t)+\sum_{k=1}^{m/2}[A_k(t)\cos(2k\theta)
+B_k(t)\sin(2k\theta)],~~~~m~~\mathrm{is~~even},
\ee
\be			\label{Gensolodd}
\int X^{m}\,w(X,\theta,t)dX=\sum_{k=0}^{(m-1)/2}\{A_k(t)\cos[(2k+1)\theta]
+B_k(t)\sin[(2k+1)\theta]\},~~~~m~~\mathrm{is~~odd},
\ee
where $A_k(t)$, $B_k(t)$ are independent on $\theta$, but dependent on time $t$.

Consistent with (\ref{Gensoleven}) or (\ref{Gensolodd}), from formula (\ref{optfromWig}) we have that
all average values $\langle X^m\rangle$ 
are the sums of terms of the form $A_{k\,m}(t)\cos^k\theta\sin^{m-k}\theta$, 
where $k=0,~1,...,m$ and $A_{k\,m}(t)$ are the constants that depend only on time. 
In other words the following conditions must be satisfied:
\be			\label{garmsolQPnW}
\int X^{m}\,w(X,\theta,t)dX=
\sum_{k=0}^{m}A_{km}(t)
\cos^k\theta\sin^{m-k}\theta,~~~m=1,~2,...,~n-2,
\ee
\be			\label{coefAkm}			
A_{km}(t)=
C_{m}^k\int q^kp^{m-k}\,
W(q,p,t)dq~dp,
\ee
where $C_{m}^k$ is a binomial coefficient.

If the potential $V(q)$ is an analytic function representing an endless 
series of powers of $q$,
then the tomogram $w(X,\theta,t)$ must satisfy an infinite number of conditions 
of the form (\ref{garmsolQPnW}), where $n=\infty$. 

Obviously, if we consider the classical case, 
then in formula (\ref{coefAkm}) the Wigner function $W(q,p,t)$ must be replaced 
with classical distribution function $W_\mathrm{cl}(q,p,t)$.

One can show that to all these conditions be satisfied, the function
$w(X,\theta,t)$ must allow to be expressed as an expansion in Hermite polynomials of the following form:
\be			\label{razlozhH}
w(X,\theta,t)=\sum_{n,\,m}\rho_{nm}(t)
\frac{e^{i\theta(m-n)}e^{-X^2}}{\sqrt{\pi2^{n+m}n!m!}}
H_n(X)H_m(X),
\ee
where for the quantum case the matrix $\rho_{nm}(t)$ is obviously a density matrix
of the quantum state in Fock basis representation, and for the classical case the set of quantities
$\rho_{nm}^\mathrm{cl}(t)$ can be expressed in terms of classical distribution function
$W_\mathrm{cl}(q,p,t)$ as follows:
\bea			\label{rhoclas}
\rho_{nm}^\mathrm{cl}(t)&=&\frac{1}{\sqrt{\pi2^{n+m}n!m!}}
\int W_\mathrm{cl}\left(\frac{q+q'}{2},p,t\right) \nonumber\\[3mm]
&&\times\exp\left(ip\,(q-q')
-\frac{q^2}{2}-\frac{q'^2}{2}\right)
H_n(q)H_m(q')\,dq~dq'~dp.
\eea
So that, for the normalization of the optical tomogram be preserved during evolution 
in accordance with the equation (\ref{r20_2}) or (\ref{eq_54}),
it is necessary and sufficient that the tomogram must be a sum of the form
(\ref{razlozhH}), which can be written for $N$-dimensional system as:
\bea			\label{razlozhHN}
w(\mathbf X,\bm\theta,t)=\sum_{{n_1...n_N}\atop{m_1...m_N}}
\rho_{n_1...n_Nm_1...m_N}(t)\frac{1}{\sqrt{\pi^N}}
\prod_\sigma^N
\frac{e^{i\theta_\sigma(m_\sigma-n_\sigma)}e^{-X_\sigma^2}}
{\sqrt{2^{n_\sigma+m_\sigma}n_\sigma!m_\sigma!}}
H_{n_\sigma}(X_\sigma)H_{m_\sigma}(X_\sigma).
\eea

Note that the property of conservation of normalization in itself, 
additionally to non-negativity,  parity, and satisfaction of Hirschmann criterion,
is also not sufficient for the function to be a tomogram. Consider the example
\be			\label{example3}
w_1(X,\theta)=\frac{e^{-X^2}}{\sqrt\pi}\left(X^4+\frac{X^2}{4}+\frac{1}{8}\right).
\ee
Despite the fact that the function $w_1(X,\theta)$ has all properties listed above,
it is not a tomogram of any quantum or classical state,
because the ``probability" of finding a quantum system in state $|0\rangle\langle0|$
is negative (equal to $-1/8$) or, if we apply to $w_1(X,\theta)$ transformation (\ref{eq_52}),
we obtain the function
\be			\label{example4}
W_1(q,p)=\frac{e^{-q^2-p^2}}{\pi}\left[(q^2+p^2)^2-\frac{3}{4}(q^2+p^2)-\frac{1}{4}\right],
\ee 
which is not a classical distribution function.

\section{\label{Section5}Conditions for conservation of normalization \\
of symplectic tomogram-like functions during evolution}
For Hamiltonians  (\ref{r18_2})
the evolution equation 
of the symplectic tomogram  $M({\mathbf X},{\bm\mu,\bm\nu},t)$ 
of the quantum system has the form \cite{ManciniFoundPhys97}
\begin{eqnarray}
\frac{\partial}{\partial t}M({\mathbf X},{\bm\mu},{\bm\nu},t)&=&
\left[\sum_{\sigma=1}^N\frac{\mu_\sigma}{m_\sigma}\frac{\partial}{\partial\nu_\sigma}\right]
M({\mathbf X},{\bm\mu},{\bm\nu},t) \nonumber \\[3mm]
&+&
\frac{2}{\hbar}
\left[\mathrm{Im}~
V\left\{-\left[\frac{\partial}{\partial X_\sigma}\right]^{-1}
\frac{\partial}{\partial\mu_\sigma}+\frac{i\nu_\sigma\hbar}{2}
\frac{\partial}{\partial X_\sigma}\right\}\right]
M({\mathbf X},{\bm\mu},{\bm\nu},t),
\label{eq_46}
\end{eqnarray}
and for classical system in the potential field $V(\mathbf q,t)$
we have \cite{OlgaJRLR97}
\begin{eqnarray}
\frac{\partial}{\partial t}M_\mathrm{cl}(\mathbf X,\bm\mu,\bm\nu,t)&=&
\left[\sum_{\sigma=1}^N\frac{\mu_\sigma}{m_\sigma}\frac{\partial}{\partial\nu_\sigma}\right]
M_\mathrm{cl}(\mathbf X,\bm\mu,\bm\nu,t) \nonumber \\[3mm]
&+&
\left[\sum_{\sigma=1}^{N}\frac{\partial}{\partial q_\sigma}
V\left\{q_\sigma\rightarrow-\left[\frac{\partial}{\partial X_\sigma}\right]^{-1}
\frac{\partial}{\partial\mu_\sigma}\right\}
\nu_\sigma\frac{\partial}{\partial X_\sigma}\right]
M_\mathrm{cl}(\mathbf X,\bm\mu,\bm\nu,t).
		\label{eq_55}
\end{eqnarray}

Making the similar calculations as in the previous section for the term
in the potential $V(q)$ with $q^3$ we can find the equation
for the first momentum of the variable $X$
\be 			\label{condit1}
\partial^2_\mu\int XM(X,\mu,\nu,t)dX=0.
\ee
The solution of this differential equation has the linear form with
respect to the variable $\mu$
\be 			\label{solutionM1}
\int XM(X,\mu,\nu,t)dX=A_1(\nu,t)+\mu B_1(\nu,t),
\ee
and taking into account the homogeneity condition (\ref{eq_3})
we find that $B_1(\nu,t)=B_1(t)$ is independent on $\nu$
and $A_1(\nu,t)=\nu A'_1(t)$, where $A'_1(t)$ is independent on $\nu$
\be 			\label{solutionM1cor}
\int XM(X,\mu,\nu,t)dX=\mu B_1(t)+\nu A'_1(t),
\ee
or from definition (\ref{eq_44})
this relation can be written as
\be			\label{garmsolQPM}
\int XM(X,\mu,\nu,t)dX=\langle\hat q\rangle\mu
+\langle\hat p\rangle\nu.
\ee
Analogously, for the term $q^4$ in the potential $V(q)$ under conditions (\ref{eq_44}) and (\ref{condit1})
we can find the condition
\be 			\label{condit2}
\partial^3_\mu\int X^2M(X,\mu,\nu,t)dX=0,
\ee
which is valid, because 
\bea			
\int X^2M(X,\mu,\nu,t)dX&=&
\left(\int q^2W(q,p,t)dq~dp\right)\mu^2
+\left(\int p^2W(q,p,t)dq~dp\right)\nu^2 \nonumber \\[3mm]
&&+2\left(\int qp\,W(q,p,t)dq~dp\right)\mu\nu.
\label{garmsolQPMW}
\eea
For the polynomial potential of the power of $q^n$ as in the cases of (\ref{condit1}) and (\ref{condit2})
we can write 
\be 			\label{conditn}
\partial^{m-1}_\mu\int X^{m-2}M(X,\mu,\nu,t)dX=0,~~~m=2,~3,...,~n,
\ee
and as in the case of (\ref{garmsolQPnW})
we have
\be			\label{garmsolQPnWM}
\int X^{m}\,M(X,\mu,\nu,t)dX=
\sum_{k=0}^{m}A_{km}(t)
\mu^k\nu^{m-k},~~~m=1,~2,...,~n-2,
\ee
where $A_{km}(t)$ are defined by formula (\ref{coefAkm}),
and so, the symplectic tomogram must be represented in the form of the sum \cite{OlgaJRLR97}
\bea
M(X,\mu,\nu,t)&=&\sum_{n,m}\frac{\rho_{nm}(t)}
{\sqrt{\pi(\mu^2+\nu^2)2^{(n+m)}n!m!}}
\frac{(\nu+i\mu)^m(\nu-i\mu)^n}{(\mu^2+\nu^2)^{(n+m)/2}}
\nonumber \\[3mm]
&&\times\exp\left(-\frac{X^2}{\mu^2+\nu^2}\right)
H_n\left(\frac{X}{\sqrt{\mu^2+\nu^2}}\right)
H_m\left(\frac{X}{\sqrt{\mu^2+\nu^2}}\right),
\label{razlMH}
\eea
which is the necessary and sufficient condition for preservation
of the normalization  during evolution of the symplectic tomogram.
For $N-$dimensional systems the latter formula has the form
\bea			
M(\mathbf X,\bm\mu,\bm\nu,t)&=&\sum_{{n_1...n_N}\atop{m_1...m_N}}
\frac{\rho_{n_1...n_N\,m_1...m_N}(t)}{\sqrt{\pi^N}}
\prod_\sigma^N
\frac{(\mu_\sigma+i\nu_\sigma)^{m_\sigma}(\mu_\sigma-i\nu_\sigma)^{n_\sigma}}
{\sqrt{2^{n_\sigma+m_\sigma}(\mu_\sigma^2+\nu_\sigma^2)^{n_\sigma+m_\sigma+1}
n_\sigma!m_\sigma!}} \nonumber \\[3mm]
&&\times
\exp\left(-\frac{X_\sigma^2}{\mu_\sigma^2+\nu_\sigma^2}\right)
H_{n_\sigma}\left(\frac{X_\sigma}{\sqrt{\mu_\sigma^2+\nu_\sigma^2}}\right)
H_{m_\sigma}\left(\frac{X_\sigma}{\sqrt{\mu_\sigma^2+\nu_\sigma^2}}\right).
\label{HermExpanN}
\eea

As in the case with the optical tomograms, conditions of conservation of normalization of type 
(\ref{conditn}), or (\ref{garmsolQPnWM}), or (\ref{razlMH}) are not sufficient for the non-negative
tomogram-like function $f(\mathbf X,\bm\mu,\bm\nu)$ obeying to (\ref{eqnormSymp}) and (\ref{eq_3})
to be a tomogram of any quantum or classical state.
Proving example it is easy to obtain from expression (\ref{example3}) applying the formula
(\ref{eq_6}):
\be			\label{example5}
M_1(X,\mu,\nu)=\frac{1}{\sqrt\pi\sqrt{\mu^2+\nu^2}}
\left(\frac{X^4}{(\mu^2+\nu^2)^2}+\frac{X^2}{4(\mu^2+\nu^2)}+\frac{1}{8}\right)
\exp\left(-\frac{X^2}{\mu^2+\nu^2}\right).
\ee
The function $M_1(X,\mu,\nu)$ is non-negative, normalized, homogeneous, and it conserves
the normalization during time evolution, but it is not a tomogram.

\section{\label{Section6}Conservation of normalizations of the Wigner function,
classical distribution function, and Husimi function
during evoluton}

In previous sections we have found that in order to as classical as quantum
dynamical equations for optical and symplectic
tomograms retain normalizations, the tomograms 
have to satisfy a set of conditions.
And only normalization of the tomogram is insufficient for 
this condition remain valid during time evolution.

On the contrary, Moyal equation for the Wigner function
\cite{Moyal1949}
\be			\label{Moyal}
\frac{\partial }{\partial _t} W({\bf q},{\bf p},t)=
-\sum_{\sigma=1}^{N} \frac{p_\sigma}{m_\sigma}
\frac{\partial}{\partial_{q_\sigma }}W({\bf q},{\bf p},t)
-2\left[\mathrm{Im}~V\left\{
{\bf q}+\frac{i}{2} \frac{\partial}{\partial {\bf p}}
\right\}\right]W({\bf q},{\bf p},t),
\ee
Liouville equation for the distribution function
\be			\label{Vlasov}
\frac{\partial }{\partial _t} W_\mathrm{cl}({\bf q},{\bf p},t)=
-\sum_{\sigma=1}^{N} \frac{p_\sigma}{m_\sigma}
\frac{\partial}{\partial_{q_\sigma }}W_\mathrm{cl}({\bf q},{\bf p},t)
-\left[\frac{\partial }{\partial {\bf q}}~V({\bf q})\right]\frac{\partial}{\partial {\bf p}}
W_\mathrm{cl}({\bf q},{\bf p},t),
\ee
and  dynamical equation for the Husimi 
function \cite{Mizrahi1986}
\bea			\label{HusimiEQ}
\frac{\partial }{\partial _t} Q({\bf q},{\bf p},t)&=&
-\sum_{\sigma=1 }^{N} \frac{1}{m_\sigma}\left[p_\sigma
\frac{\partial}{\partial_{q_\sigma }}
+\frac{1}{2}\frac{\partial}{\partial_{p_\sigma }}
\frac{\partial}{\partial_{q_\sigma }}\right]
Q({\bf q},{\bf p},t) \nonumber\\[3mm]
&&-2\left[\mathrm{Im}~V\left\{
{\bf q}+\frac{1}{2}\frac{\partial }{\partial {\bf q}}
+\frac{i}{2} \frac{\partial}{\partial {\bf p}}
\right\}\right]Q({\bf q},{\bf p},t),
\eea
retain normalization for any normalized and quickly decaying functions at $q_i$, $p_j$ 
tending to infinity.
As in the previous parts let's approximate the potential $V(\bf q)$ 
with a polynomial and integrate both sides of the dynamical equations 
(\ref{Moyal}) -- (\ref{HusimiEQ}) over the hyperspace ${\bf q}\times{\bf p}$. 
Easy to see that for any functions $W({\bf q},{\bf p},t)$, 
$W_\mathrm{cl}({\bf q},{\bf p},t)$ and $Q({\bf q},{\bf p},t)$  from the class considered
the right sides of equations (\ref{Moyal}) -- (\ref{HusimiEQ})
are becoming identically equal to zero.

Thus, Moyal equation, Liouville equation and dynamical equation for the Husimi function 
retain normalizations of their solutions without any additional conditions, 
like (\ref{Gendifeven}, \ref{Gendifodd}) for optical
or (\ref{conditn})  for symplectic tomograms and that is a principal 
difference of equations (\ref{Moyal}) -- (\ref{HusimiEQ}) from corresponding 
equations (\ref{r20_2}, \ref{eq_54}, \ref{eq_46}, \ref{eq_55}) for tomograms.

\section{\label{Section7}Conclusion}

In summary we point out the main results of our paper.
We discussed the properties of optical and symplectic tomograms. 

It was demonstrated that optical tomograms could be associated with a 
unitary representation of the Weyl-Heisenberg group as well as symplectic 
tomograms. This fact has allowed to formulate an autonomous criterion for optical
tomogram-like function to be a tomogram of quantum or classical state
based on positivity properties of the function on the group
determined with the help of the latter tomogram-like function.

We shown that for the arbitrary potential the tomograms must satisfy  a set 
of relations to be normalized during evolution.
We found that all moments of homodyne variable $X$ 
as for optical as for symplectic tomography
must satisfy a set of specific linear differential equations.
These equations were obtained explicitly and their general 
solutions were presented.

We also illustrated that Moyal equation, Liouville equation, and dynamical equation
for the Husimi function retain the normalization for any normalized and decreasing
rapidly at infinity initial conditions, unlike tomographic dynamical equations.

The relations allowing to conserve the normalization of the tomogram
during time evolution determine a narrow class of functions, which can 
evolve without dramatic growth maintaining the physical sense.

\end{document}